\documentclass[twocolumn,3p]{elsarticle_mod}

\usepackage{graphicx}
\usepackage{booktabs}
\usepackage{amsmath, amssymb}
\usepackage{braket}
\usepackage{hyperref}

\title{Potential of a longitudinally focusing insertion for a storage ring X-ray FEL.}

\author[ia]{I. ̃Agapov\corref{cor1}}

\address[ia]{European XFEL GmbH, Hamburg, Germany}
\cortext[cor1]{ilya.agapov@xfel.eu}

\begin{document}



\begin{abstract}
In present work we investigate the potential of a longitudinally focusing device to compress bunches passing an undulator for a synchrotron storage ring.
If integrated into a storage ring similar to PETRAIII such device could potentially produce continuous $\sim$1ps pulses of
photons in the $nm$ range with peak pulse powers of tens of GW. Even without operating in FEL saturation mode the longitudinal focusing can provide
means to increase the brightness and shorten the photon pulse length. 
\end{abstract}

\maketitle

\section{Introduction}

\begin{figure}[h!]
 \centering
 \includegraphics[width=80mm, height=50mm]{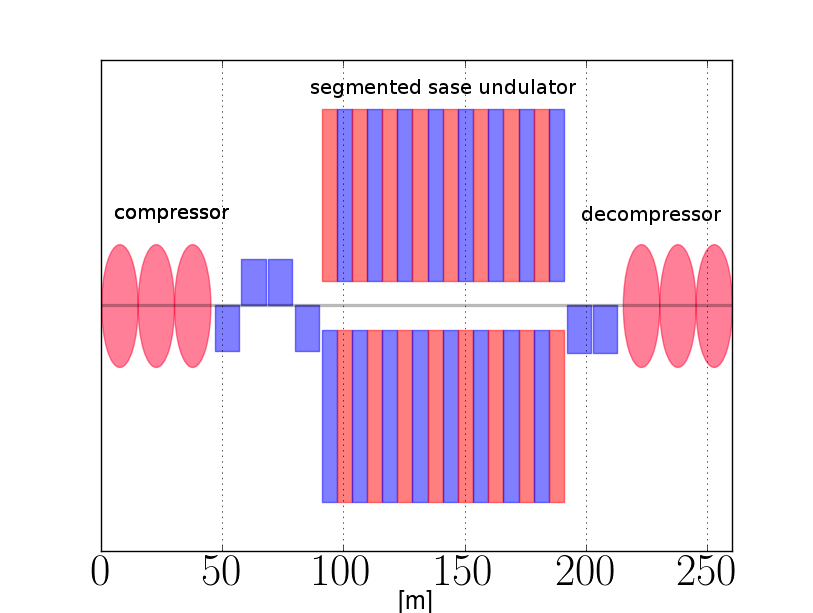}
 \caption{ Insertion device layout. Going from left to right, the beam passes an RF module, a dispersive section (chicane or arc), a number of 
 undulators, a disperive section and finally another RF module.}
 \label{fig:layout}
\end{figure}

Low gain FELs with wavelength down to $\sim200nm$ (6eV) have been in operation at storage rings using optical cavities(refs).
Short wavelength FELs presently use linacs as drivers since they provide necessary electron beam quality. 
X-ray FELs such as LCLS, European XFEL or SWISS FEL are now in operation or under construction worldwide. They use linacs as drivers 
to assure beam qualities necessary for a SASE process at those wavelengthes. For a typical wavelength (1KeV-30KeV) the European XFEL 
requires emittances below $10^6$, energy spreads $\sim 1 MeV$
and peak currents ov several $kA$ at electron beam energies up to $17.5GeV$. The saturation length (for basic definitions in the FEL theory see e.g. \cite{Saldin}) roughly defines the minimum practically sensible 
undulator length. At European XFEL, for the shortest wavelength, achieved with the maximum electron beam energy, 
the saturation length can be a hundred meters, but for 
soft X-rays it can be as short as 30 meters depending on the wavelength and electron beam parameters. This makes it in principle possible to
fit such an undulator into a storage ring. 
Beam parameters in latest generation light sources such as PETRAIII (see Tab. \ref{tab:p3}) are such that for 
UV photons the beam quality is not far removed from that required for an FEL. For shorter wavelength saturation length becomes larger and the possibility of
using the stored beam for SASE FEL directly becomes limited.  The interest in shorter wavelength storage-ring based FELs has recently been growing since they could combine extreme
peak brightness and coherence of an FEL with continuous operation and lower power consumption of a storage ring (see e.g. \cite{huang} and references therein)
An insertion device with longitudinal focusing (of crab cavity type, discussed e.g. in  \cite{zholents}) would consist of a compression section, sase undulator, 
and a decompression section. The rest of the ring could be passed with the usual bunch length. 
A design sketch is presented in Fig. \ref{fig:layout}. It could be used as an insertion or as a bypass subject to space availability.
In the following some simulation studies for the possibility of integrating such an insertion into PETRAIII are presented. All calculations are performed with xcode \cite{xcode}.

\begin{table}[h!]\label{tab:p3}
\begin{center}
\begin{tabular}{ll}
 \hline 
 {\bf Parameter } & {\bf Value } \\
 \hline 
 Beam energy & 6 GeV \\
 Circumference & 2304 m \\
 Emittance $\varepsilon_x,\varepsilon_y$ & $10^{-9}$, $10^{-11}$ \\
 Energy spread  & $10^{-3}$ (6MeV) \\
 Bunch charge  & 20 nC \\
 Bunch length  & 44ps or 13mm \\
 Peak current  & 170A \\
 Longitudinal damping time  & 10msec \\
 \hline 
\end{tabular} 
\caption{PETRAIII beam parameters \cite{p3}, assuming high bunch charge operation mode with 40 bunches}
\end{center}
\end{table}

\section{Possibility of an FEL insertion device at PETRAIII}

\paragraph{Longitudinal phase space focusing}

In a linac-based FEL bunch compression is a key factor, but can allow for certain beam distortion as long as it preserves the lasing bunch core.
For a multiturn operation the margin for such distortions is much thinner. Space charge and Coherent Synchrotron Radiation (CSR) \cite{csr} effects play much smaller role for
longer bunches and higher energies, so cleaner compression and decompression can be in principle expected than for a typical
linac FEL. The requirement is that no beam instabilities and distortions should appear on the time scale faster than the longitudinal damping time which is about 1000 turns for PETRAIII. 
Neglecting collective interactions, the longitudinal phase space map for the insertion is 

\begin{equation}\label{the_equation}
M = M_{RF2} \cdot M_{C2} \cdot M_{C1} \cdot M_{RF1}
\end{equation}

where the dispersive sections maps are given by matrices

\begin{equation}
 M_{C1,C2} = 
 \begin{pmatrix}
        1 & R_{56}^{(1,2)}\\
        0 & 1\\
  \end{pmatrix}
\end{equation}

and the RF cavity maps are

\begin{equation}
 M_{RF1, RF2} :
 \begin{pmatrix}
        t \\
        p \\
 \end{pmatrix}
 \to 
 \begin{pmatrix}
        t \\
        p + V^{(1,2)}  \sin(f_{RF} \cdot t) \\
 \end{pmatrix}
\end{equation}

\begin{figure}[h!]
 \centering
 \includegraphics[width=75mm, height=50mm]{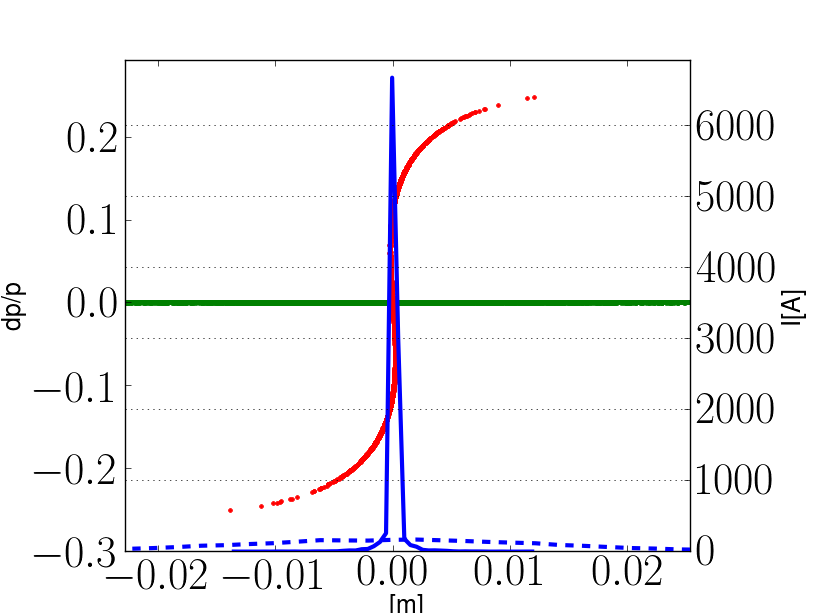}
 \caption{Simulated longitudinal phase space at the entrance (green), after the bunch compressor and at the exit
 (identical to the entrance phase space) of the insertion device,
  with 800MHz cavity.Current profiles before (blue dashed line) and after (solid blue line) are shown.}
 \label{fig:compr1}
\end{figure}

\begin{figure}[h!]
 \centering
 \includegraphics[width=80mm, height=50mm]{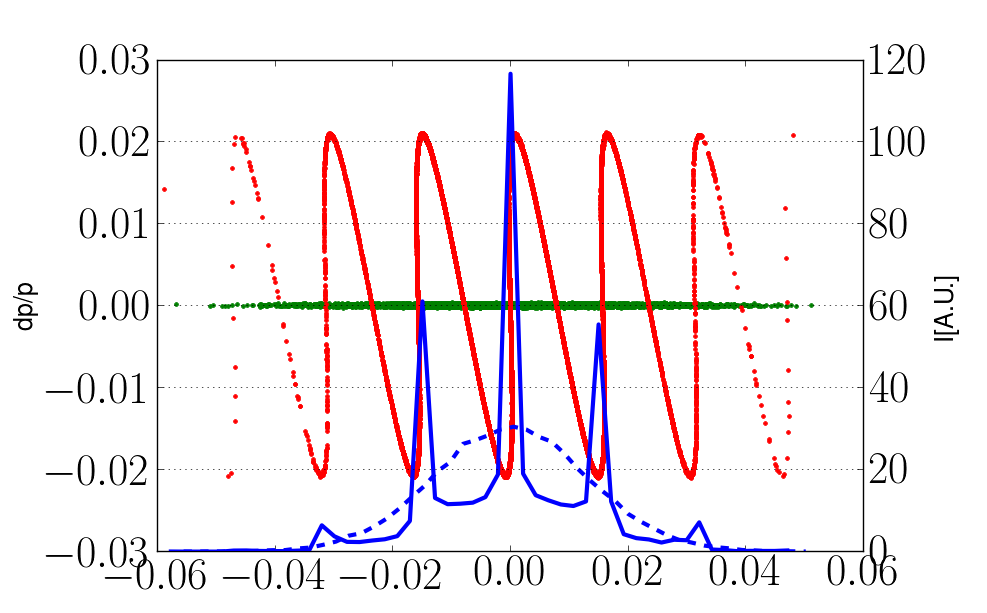}
 \caption{Simulated longitudinal phase space at the entrance (green) and after the bunch compressor (red) with 12GHz cavity. 
 Current profiles before (blue dashed line) and after (solid blue line) are shown.}
 \label{fig:compr2}
\end{figure}

Here $t$ and $p$ are longitudinal coordinates usually measured in meters and relative energy units, $R_{56}$ is the standard notation
for dispersive time delay and $V^{(1,2)}$ are total RF voltages for the compressor and the decompressor. One easily checks that by choosing
$R_{56}^{(1)} = - R_{56}^{(2)}$ and $V^{(1)} = - V^{(2)}$ the whole transfer map reduces to unity. So when collective 
self-interactions and diffusion are neglected the insertion has no theoretical footprint on longitudinal beam dynamics.
The beam compressor parameters which were used in calculations corresponding to PETRAIII beam parameters are shown in Tab. \ref{tab:BC}.
800MHz cavity was chosen for demonstration since it has the wavelength longer than the electron beam size.  1.3GHz cavity of European XFEL
type could also be used, but half its wavelength is shorter than the electron bunch length and the bunch tails will not be compressed fully (see Fig.
\ref{fig:compr1}, \ref{fig:compr2}). The system length with undulator is of the order of 200m. 
No optimization has been performed so far wrt. length, RF power, potentaial for using storage ring arcs etc., so this number could be sufficiently 
improved.

\begin{table}[h!]\label{table:und}
\begin{center}
\begin{tabular}{ll}
 \hline 
 {\bf Parameter } & {\bf Value }  \\
 \hline 
 Period & $l_w = 0.02m$\\
 Field  & $K = 0-10.0$ \\
 Radiation wavelength at 6GeV & 300eV-17KeV \\
 Section length  & 0.9m \\
 Optics  & FODO with $\beta=4m$ \\
 \hline 
\end{tabular} 
\caption{Possible parameters of a 20mm soft x-ray undulator. }
\end{center}
\end{table}

\begin{table}[h!]\label{table:und2}
\begin{center}
\begin{tabular}{ll}
 \hline 
 {\bf Parameter } & {\bf Value }  \\
 \hline 
 Period & $l_w = 0.068m$\\
 Field  & $K = 0-10.0$ \\
 Radiation wavelength at 6GeV & 100eV-5KeV \\
 Section length  & 0.9m \\
 Optics  & FODO with $\beta=4m$ \\
 \hline 
\end{tabular} 
\caption{Parameters of a 68mm soft x-ray undulator of European XFEL type (for 6GeV electron beam). }
\end{center}
\end{table}

\begin{table}[h!]\label{tab:BC}
\begin{center}
\begin{tabular}{ll}
 \hline 
 {\bf Parameter } & {\bf Value }\\
 \hline 
 RF frequency & 800MHz \\
 Cavity length (at 35MV/m gradient) & 45m x2\\
 Chicane length & 20-30m \\
 Dipole fields & ~0.2T \\
 $R_{56}$ compressor & -0.15m \\
 $R_{56}$ decompressor & 0.15m \\
 \hline 
\end{tabular} 
\caption{Parameters of the bunch compressor and decompressor used in simulations with 20x compression for nominal bunch length.}
\end{center}
\end{table}

An obvous drawback of the system is the large amount of RF power required. Since the longitudinal focusing is proportional to $V\cdot f_{RF}$
higher RF frequency will result in shortening the section length. With 12GHz cavities the length can be reduced to a few meters only. 
However the bunch length at PETRAIII is longer than the RF wavelength for frequencies about $\gtrsim 1GHz$. One can still achieve compression
in that case, but it would result in a specific current shape \ref{fig:compr2}. If low momentum compaction operation is assumed where
the bunch length is shortened by a factor of say, 10, such high frequency cavities could be employed efficiently (see Tab. \ref{tab:BC2}).

\begin{table}[h!]\label{tab:BC2}
\begin{center}
\begin{tabular}{ll}
 \hline 
 {\bf Parameter } & {\bf Value }\\
 \hline 
 RF frequency & 3.9GHz \\
 Cavity length (at 35MV/m gradient) & 9m x2\\
 Chicane length & $\sim$20m \\
 Dipole fields & $\sim$0.2T \\
 $R_{56}$ compressor & -0.15m \\
 $R_{56}$ decompressor & 0.15m \\
 \hline 
\end{tabular} 
\caption{Parameters of the bunch compressor and decompressor with 50x compression for short bunch operation.}
\end{center}
\end{table}

\paragraph{Undulator}

The radiation wavelength is 

\begin{equation}
\lambda_r = \frac{l_w}{2  \gamma^2} \left(1 + \frac{K^2}{2} \right)
\end{equation}

And the Pierce parameter is 

\begin{equation}
\rho = \frac{1}{\gamma}  \left( \left( \frac{K  A_{JJ}  l_w} { 8  \pi  \sigma_b} \right)^2  \frac{I}{I_A} \right) ^{\frac{1}{3}}
\quad I_A = 17kA
\end{equation}

gain length
\begin{equation}
L_{G} = l_w / (4 \pi  \sqrt{3} \rho )
\end{equation}

\begin{figure}[h!]
 \centering
 \includegraphics[width=80mm, height=50mm]{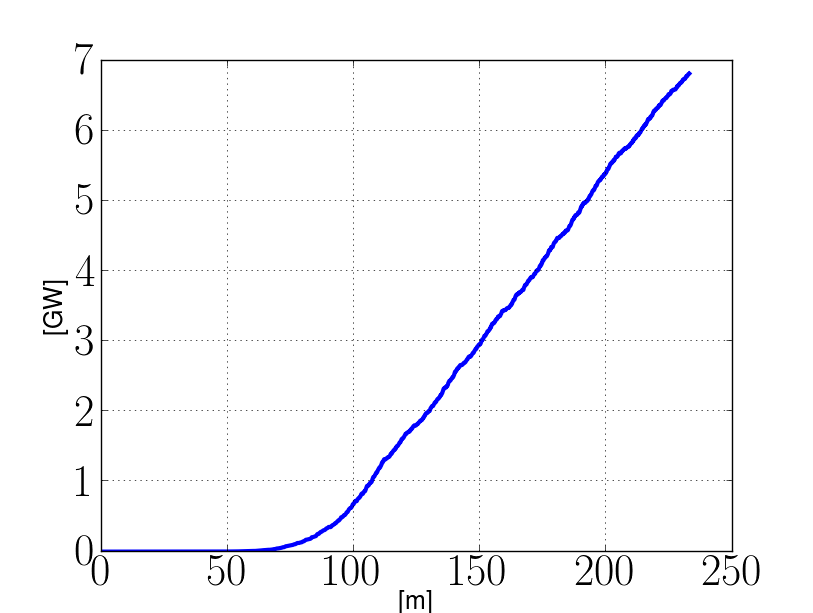}
 \caption{Steady state SASE simulations for $E_{\gamma}=1266eV$, K=5.0, 50x compression}
 \label{fig:sat1}
\end{figure}

\begin{figure}[h!]
 \centering
 \includegraphics[width=80mm, height=50mm]{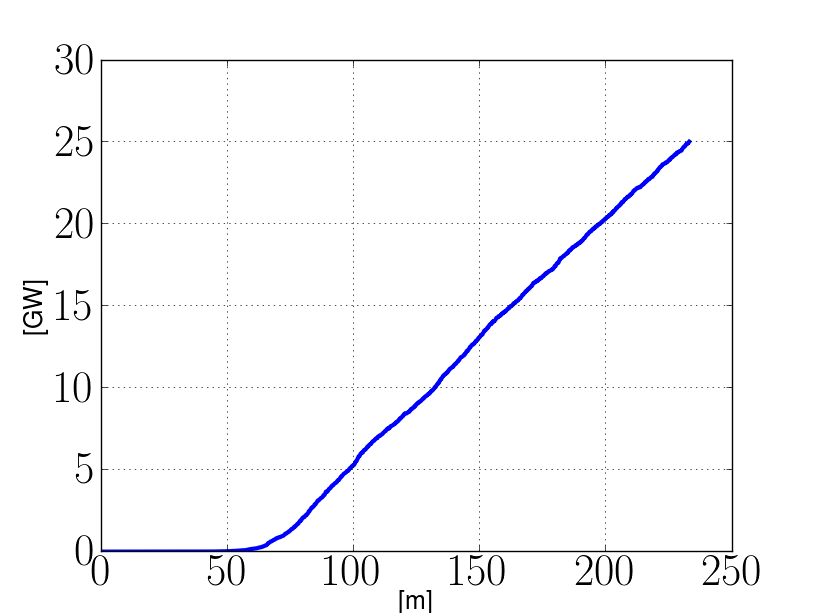}
 \caption{Steady state SASE simulations for $E_{\gamma}=335eV$, K=10.0, 50x compression}
 \label{fig:sat2}
\end{figure}

\begin{figure}[h!]
 \centering
 \includegraphics[width=80mm, height=50mm]{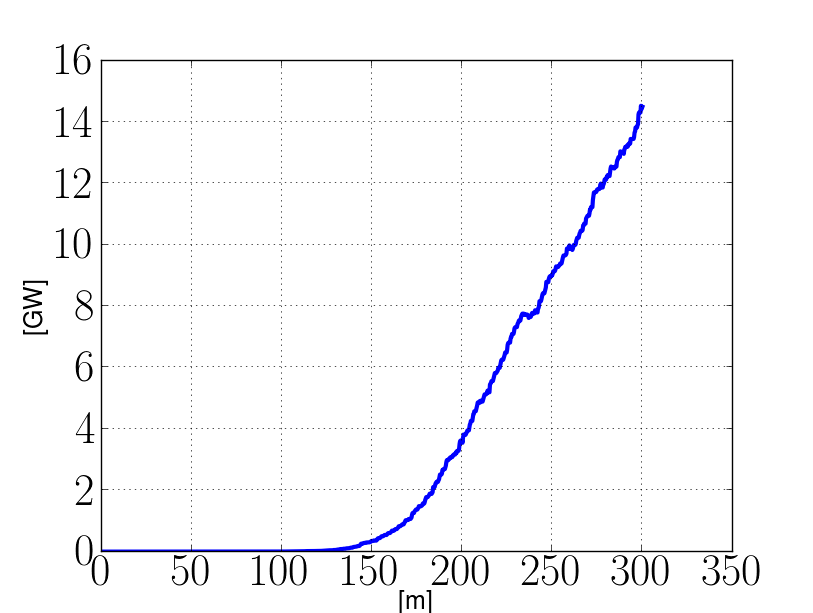}
 \caption{Steady state SASE simulations for $E_{\gamma}=100eV$, $l_w=0.068$, K=10.0, 50x compression}
 \label{fig:sat3}
\end{figure}

So to increase the energy reach of a SASE FEL into the harder part of the spectrum one chooses possibly short undulator period. If using the FODO optics, decreasing the beam size is possible down to $\sqrt{2\epsilon L}$
where $L$ is the period length. E.g. 4m FODO optics will require short undulators of $\sim1m$ length with quadrupoles between them. 
Power estimates assuming 0.02mm and 0.068mm period undulators are presented in  
Fig. \ref{fig:sat1}-\ref{fig:sat3}.

\paragraph{Influence on beam dynamics}

The feasibility question from the beam dynamics point of view reduces to the turn-to-turn preservation of beam qualities. 
Synchrotron radiation induces bunch diffusion, mostly in the longitudinal phase space. 
Such diffusion is in principle a limiting factor even for linac-based FEL performance, however for the device in question the diffusion
footprint is similar to that of standard ring insertion devices and is not in principle a limiting factor. Moreover, the FEL undulators can potentially be used in place 
of damping wigglers for reducing the emittance.
A major limiting factor in FELs is the coherent synchrotron radiation \cite{csr}.  In the design discussed the bunch length (~1ps) is 5-6 orders of magnitude longer than the critical
wavelength of the bending magnet radiation (~2-20KeV for 0.1-1T dipoles), whereas CSR manifests itself when the bunch length is comparable to the wavelength of the radiation
emitted. Estimates based on taking into account the power enhancement factor

\begin{equation}
g(\lambda) = N \left| \int_{0\infty}^{\infty} n(z) \exp(2 \pi i z / \lambda ) dz \right|^2
\end{equation}

where $N$ is the number of particles in the bunch, $n(z)$ the bunch form factor and $\lambda$ the radiation wavelength, show negligible effect on the total emitted power of sycnchrotron radiation.
Another SASE undulator effect on the beam is microbunching which is washed away in dispersive sections and should not present a problem.

\section{CONCLUSION AND OUTSTANDING R\&D NEEDS}

\begin{figure}[h!]
 \centering
 \includegraphics[width=80mm, height=50mm]{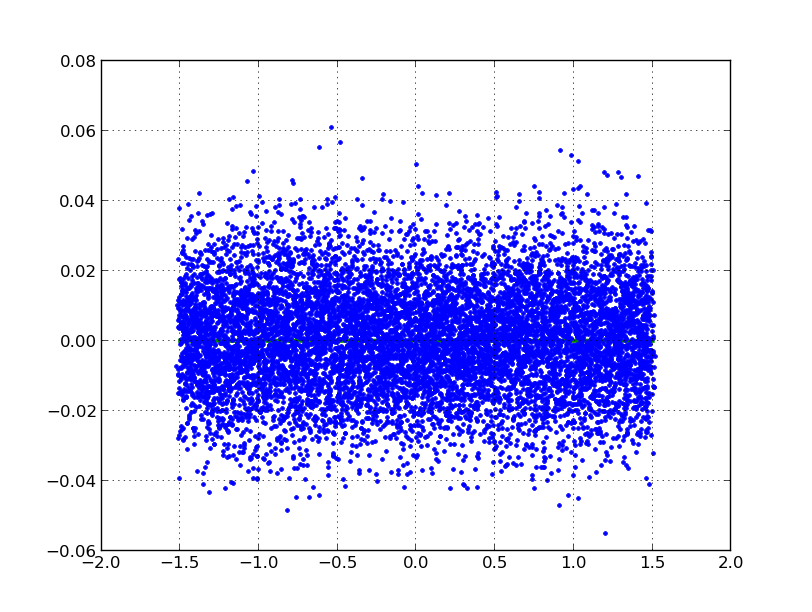}
 \caption{Effect of 1000 iterations of a map of type \ref{eq:the_equation} on a distribution with no 
 momentum spread, with 1\% random noise, normalized coordinates. }
 \label{fig:noise}
\end{figure}

\begin{figure}[h!]
 \centering
 \includegraphics[width=80mm, height=50mm]{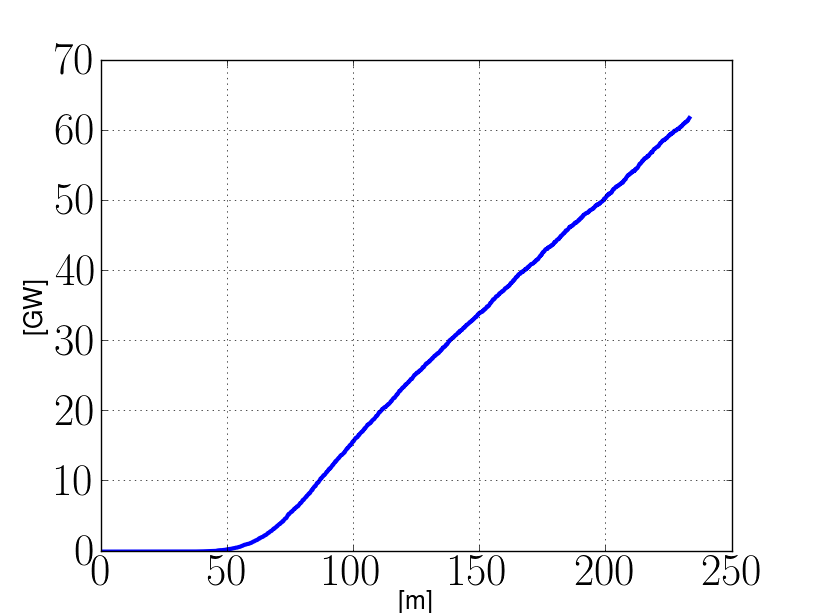}
 \caption{Steady state SASE simulations for $E_{\gamma}=335eV$, K=10, 50x compression 
 assuming $\beta=4m$ optics equalizing vertical and horizontal emittance is possible.}
 \label{fig:sat4}
\end{figure}

\begin{figure}[h!]
 \centering
 \includegraphics[width=80mm, height=50mm]{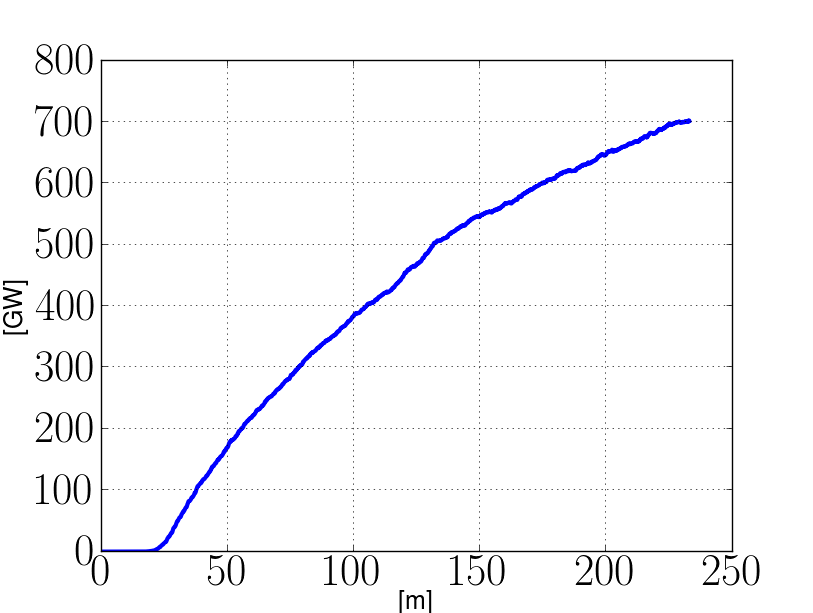}
 \caption{Steady state SASE simulations for $E_{\gamma}=335eV$, K=10, 20x compression 
 assuming $\beta=4m$ optics equalizing vertical and horizontal emittance and bunch shortening down to 4ps is possible.}
 \label{fig:sat5}
\end{figure}

Following issues need to be further addressed.

\begin{enumerate}[i.]

\item Although CSR should not be a limiting factor, possibilities for other instabilities need investigation.

\item Another possible effect is the longitudinal phase space dilution due to nonlinearity of space rotation during
compression and decompression in combinanion with diffusion or mismatching of RF phases, voltages, or $R_{56}$. 
A calculation for 1000 iterations assuming 1\% random perurbation to the one-pass map (Figure \ref{fig:noise}) suggests that the effect should notation
play a role if the compressor and decompressor pair is tuned properly.

\item Electron optics design has to be performed to match the existing beam transport. In this work power estimates were done 
based on standard FODO optics with $\left<\beta\right> = 4m$. Due to large difference in the vertical and horizontal emittances,
special optics producing round beams will improve the performance. This could be achieved by emittance exchange bewteen the vertical and 
the horizontal (see Fig. \ref{fig:sat4}) plane or possibly by an assymetric low-$\beta_x$ optics. The energy spread after the first 
focusing RF cavity is significant (5\%). This will require very large momentum acceptance optics in the insertion. 
All these questions require separate R\&D.

\end{enumerate}

With proper optimization of the layout for the X-ray wavelength of interest to experiments, 
and assuming short bunch operation ($\sim$4ps) of a storage ring,  
operation of $\sim$100m long insertion device with longitudinal focusing as an FEL
in the present or next \cite{usr} generation of light sources producing ps photon pulses of high peak power seems theoretically possible.

\section{Acknowledgements}
The author is thankful to G. Geloni and R. Wanzenberg for useful discussions.

\end{document}